\documentstyle[aps,twocolumn,floats,prd,psfig]{revtex}

\newcommand{\ApJL}{Astrophys. J. Lett.}
\newcommand{\ApJ}{Astrophys. J.}

\newcommand{\PRD}{Phys. Rev. D}

\newcommand{\sz}{{\rm SZ}}    
\newcommand{\ps}{{\rm PS}}    
\newcommand{\lss}{{\rm LSS}}    

\begin{document}
\twocolumn[\hsize\textwidth\columnwidth\hsize\csname
@twocolumnfalse\endcsname

\title{Small Angular Scale CMB Anisotropies from CBI and BIMA experiments: 
Early Universe or Local Structures?}
\author{Asantha Cooray$^1$ and Alessandro Melchiorri$^2$}
\address{$^1$Theoretical Astrophysics, California Institute of Technology, 
Pasadena, CA 91125. E-mail: asante@caltech.edu\\
$^2$Theoretical Astrophysics, University of Oxford, Oxford, OX 3RH. E-mail:melch@astro.ox.ac.uk}


\maketitle

\begin{abstract}
The advent of high resolution cosmic microwave background (CMB) experiments 
now allows studies on the temperature fluctuations at scales corresponding to 
few arcminutes and below. 
Though the reported excess power at $\ell \sim 2000 - 6000$ by CBI and BIMA 
is roughly consistent with a secondary contribution resulting from the 
Sunyaev-Zeldovich effect, this requires a higher normalization for the 
matter power spectrum than measured by other means.
In addition to a local red-shift contribution, another strong possibility for 
anisotropies at very small scales involve non-standard aspects of 
inflationary models.  To distinguish between contributions from
early universe and local structures, including a potential point source 
contribution, and to understand the extent to which structures at 
low red-shifts contribute to small scale temperature anisotropies, it may
be necessary to perform a combined study involving CMB and 
the large scale structure. We suggest a cross-correlation of the temperature 
data with a map of the large scale structure, such as the galaxy distribution.
For next generation small angular scale CMB experiments,  
multi-frequency observations may be a necessary aspect to allow
an additional possibility to distinguish between these different scenarios.
\end{abstract}
\vskip 0.5truecm
]



\section{Introduction}

The high resolution, and high signal-to-noise, 
cosmic microwave background (CMB) experiments like 
Boomerang \cite{netterfield}, DASI \cite{halverson},
MAXIMA \cite{lee}, and, more recently, VSA \cite{scott}
have revolutionized the study of temperature anisotropies 
at arcminute scales.  The multiple peaks observed by these experiments in the
angular power spectrum of the CMB anisotropies have provided
a strong evidence for inflationary models with adiabatic initial 
conditions for structure formation.

The recent results from the CBI interferometer \cite{pearson} confirms 
the presence
of multiple peaks in the angular power spectrum with an unambiguous 
detection of the damping tail as expected. 
At smaller angular scales beyond the damping tail, however,
the CBI experiment in the 'deep field' configuration 
\cite{mason} and the BIMA array \cite{bima} have both reported the 
presence of an 'excess' in temperature power at a level of about 
$\sim 500 \mu K^2$ with a detection confidence at the level of 
$\sim 3 \sigma$ and above.

\begin{figure*}[t]
\centerline{
\psfig{file=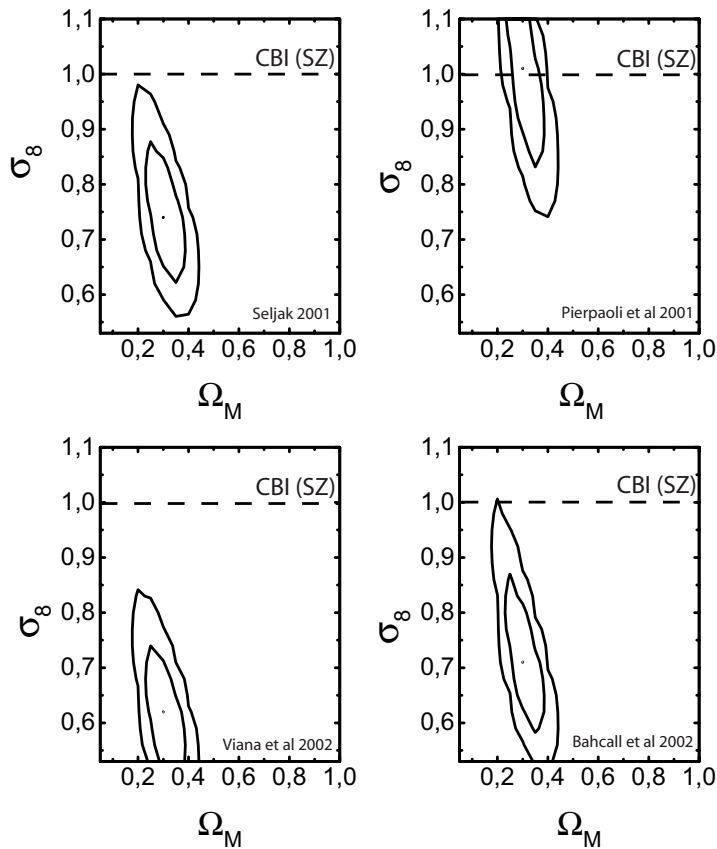,width=3.8in}}
\caption{Recent results in the $\sigma_8-\Omega_M$ plane and the
agreement with the CBI result under the assumption of a
pure SZ contribution. The analysis are from Seljak 2001, 
Pierpaoli et al. 2001, Viana et al. 2002, and Bahcall et al 2002,
 assuming a gaussian prior 
of $\Omega_M=0.31 \pm 0.05$ from the CMB+SN-Ia analysis of 
Sievers et al. 2002.}
\end{figure*}

\begin{figure*}[t]
\centerline{
\psfig{file=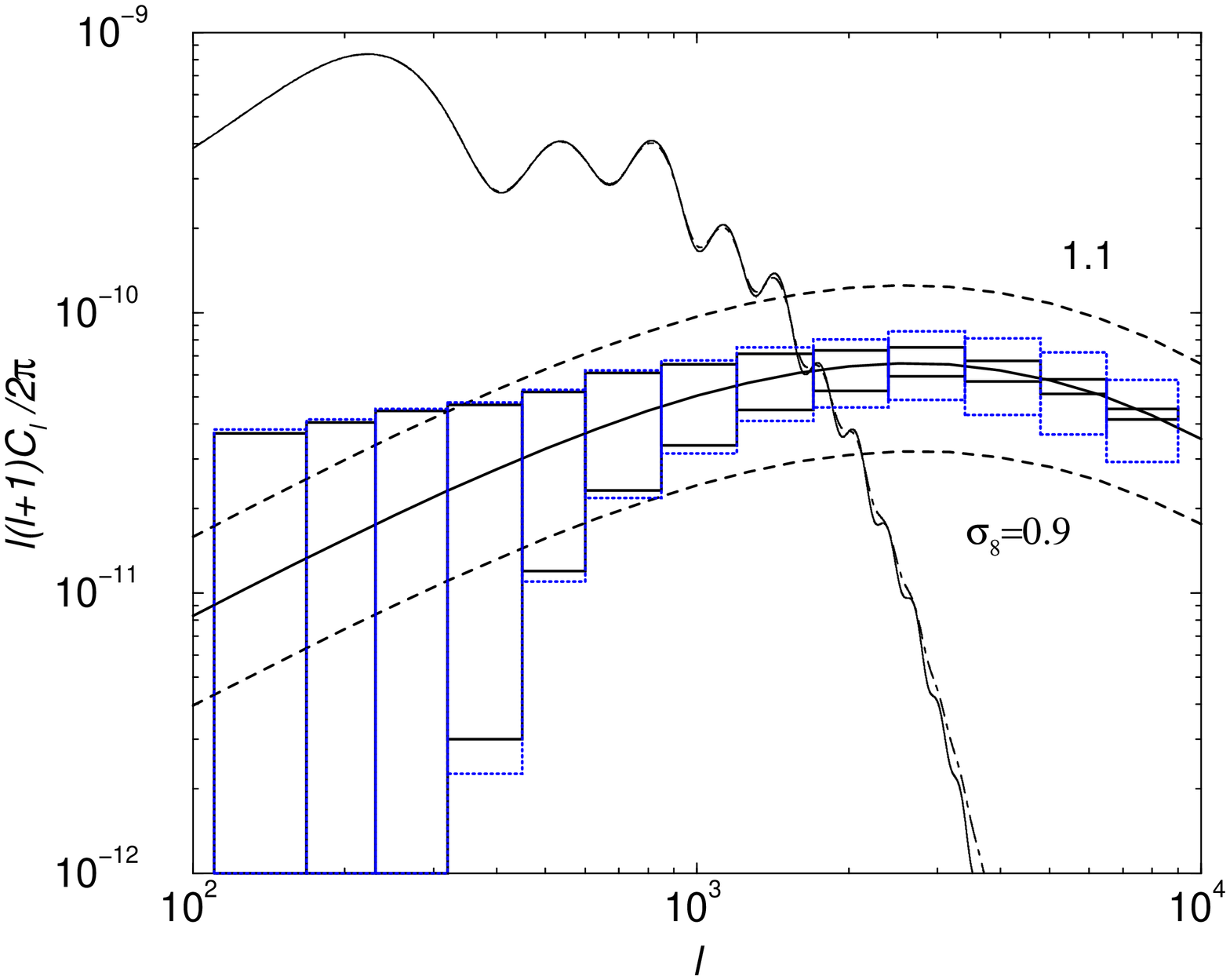,width=3.8in,angle=-90}
\psfig{file=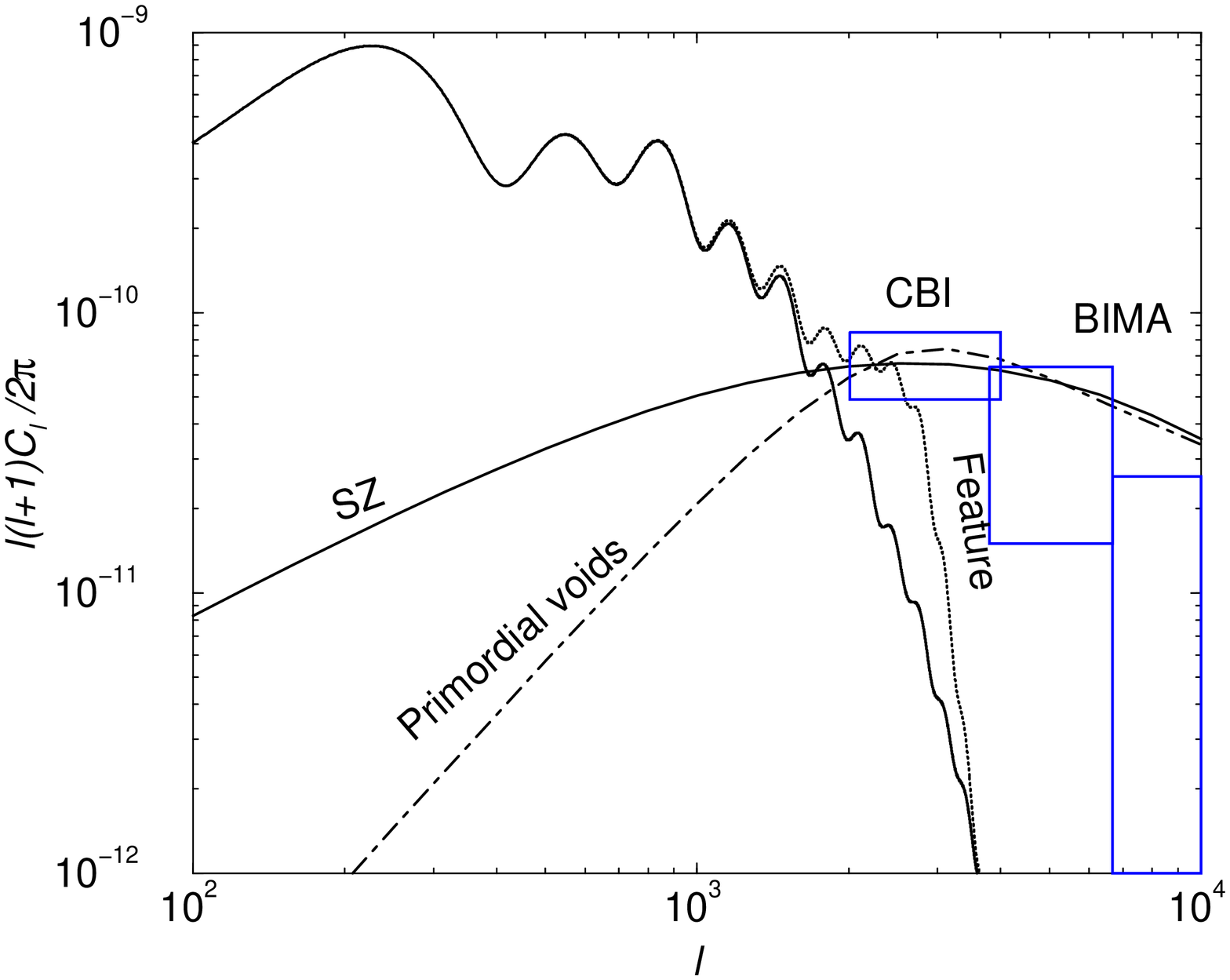,width=3.8in,angle=-90}
}
\caption{{\it Left:} The thermal SZ power spectrum. 
The three curves show the variation in the SZ contribution due to a 
change in the
normalization about $\sigma_8=1.0$. Due to the highly non-linear 
behavior, the SZ thermal contribution is strongly dependent on the
normalization of the matter power spectrum. The two sets of error bars 
show the highly non-Gaussian behavior of the clusters that
contribute: the small error bars in solid are the ones expected 
under a Gaussian description, while the dotted errors bars show the
total errors including the covariance due to non-Gaussianities. 
For illustration, we have assumed a no instrumental noise survey of
1 sqr. degrees. 
In addition to increasing the errors by a factor of a few, 
non-Gaussianities also correlate the arcminute scale band power
estimates at the 50\% to 90\% level. {\it Right:} 
Possible contributions to small scale power from effects related to the early
universe. These include a contribution from a feature in the 
primordial spectrum of fluctuations (dotted line),a distribution of 
primordial voids at the last scattering surface (dot-dashed line). 
For comparison, we also show  the
SZ contribution and the standard prediction for anisotropy power spectrum.}
\label{fig:sz}
\end{figure*}

The presence of this excess is not consistent with
the predicted damping of the primordial (primary) anisotropies 
\cite{mason} but can in principle be explained as due to a secondary 
effect \cite{bond,selkom}. At angular scales corresponding 
to projected galaxy clusters extents of order few arcminutes and below,
the Sunyave-Zel'dovich effect (SZ; \cite{SunZel80}), in fact, produces 
a now well-known contribution.  The SZ effect has now been directly imaged 
towards massive galaxy clusters \cite{Caretal96}, 
where temperature of the scattering medium  can reach as high as
10 keV producing temperature changes in the CMB of order 1 mK at
Rayleigh-Jeans (RJ) wavelengths and whose presence is a priori known 
based on optical data. 
These, and other unresolved, clusters contribute to the dominant anisotropy 
contribution at arcminute scales.

\begin{figure*}[t]
\centerline{
\psfig{file=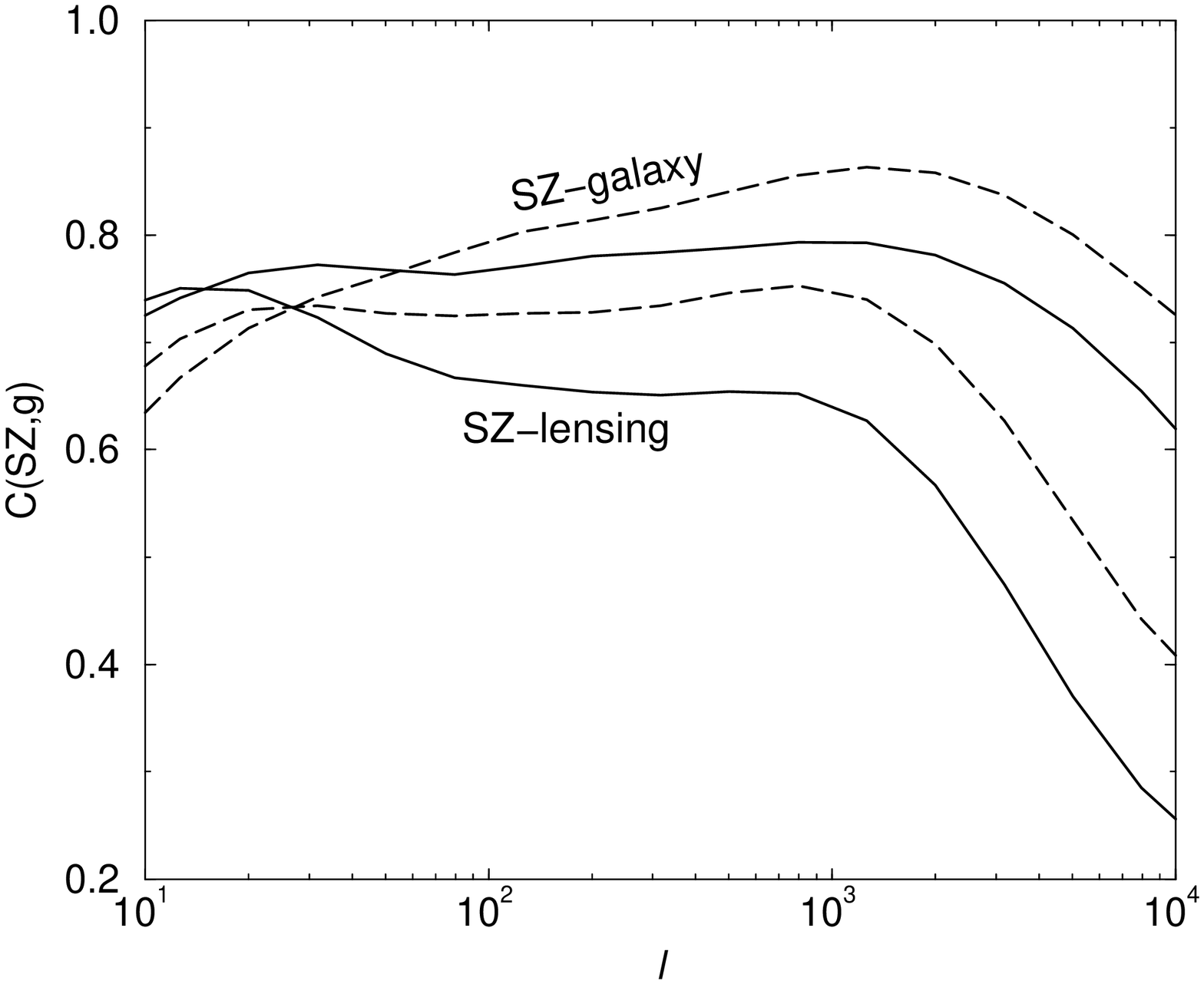,width=3.8in,angle=-90}
\psfig{file=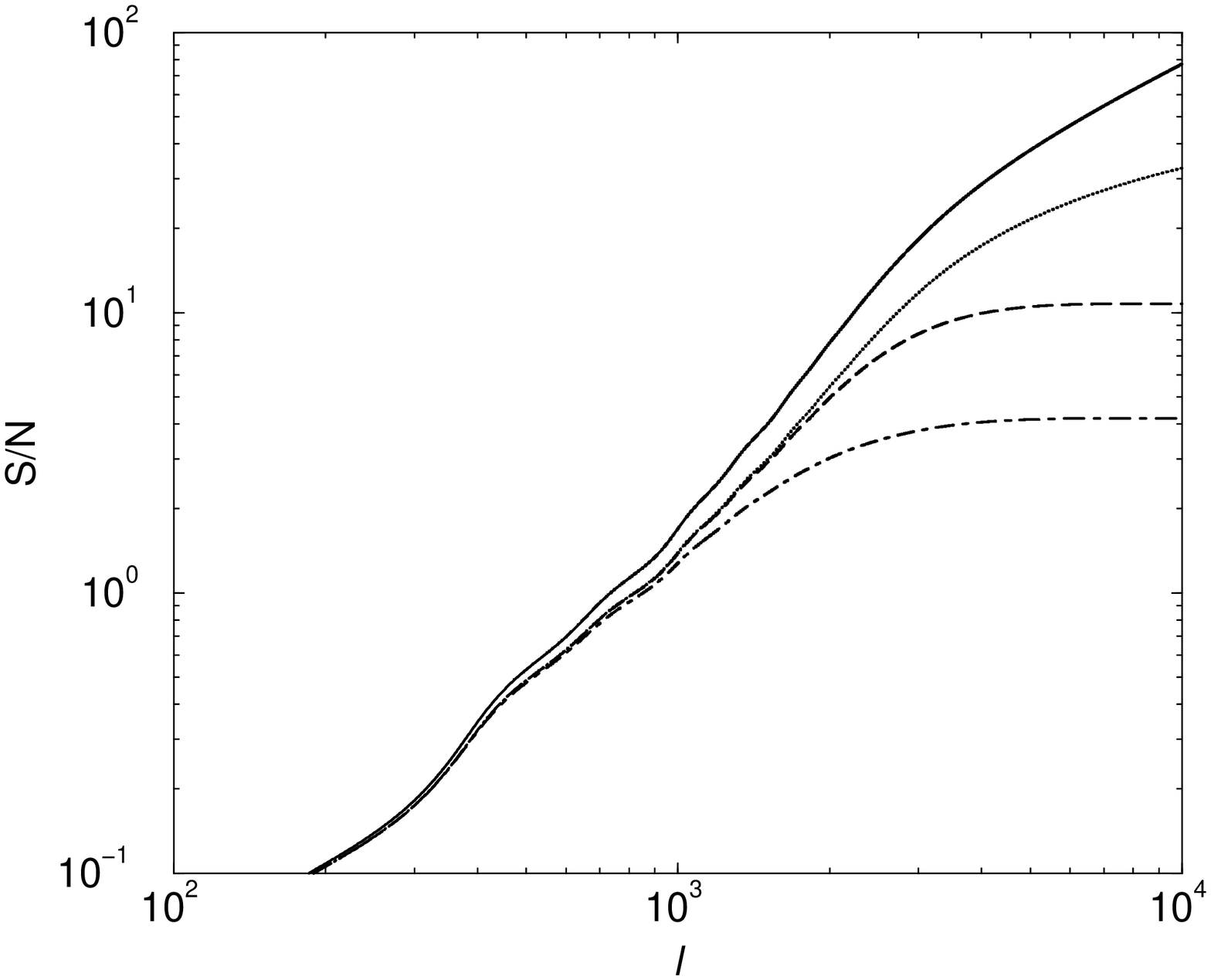,width=3.8in,angle=-90}
}
\caption{{\it Left:} The SZ-large scale structure correlation coefficients. 
The curves are for large scale structure tracers involving
galaxies, with a median redshift of 0.5 (dashed bottom) 
and 1.0 (dashed top) and weak lensing, with background sources
at redshifts of 1.0 (solid bottom) and 3.0 (solid top). 
In general, large scale structure tracers are correlated with the SZ effect
at arcminute scales with correlation coefficients of order $\sim$ 0.6, 
when median redshift involved is of order 1. {\it Right:} The
cumulative signal-to-noise ratio for the detection of SZ-galaxy 
cross-correlation. We assume a survey of
10 degrees$^2$ and a year of observations. 
The solid line is the maximum with no instrumental noise or shot-noise 
contribution, while the dotted line
is the signal-to-noise ratio with a galaxy surface density of 
$10^9$ sr$^{-1}$, the dashed line is the signal-to-noise
ratio with an additional instrumental noise for the small scale 
CMB experiment with a sensitivity of 25 $\mu$K $\sqrt{\rm sec}$ and
a beam of 2 arcmins (FWHM) and the dot-dashed line is the ratio when 
the sensitivity is 100 $\mu$K $\sqrt{\rm sec}$.}
\label{fig:corr}
\end{figure*}

Even if the detection by the CBI and BIMA experiments 
marks an important milestone, however, it is still not clear if the excess
can be easily explained by the SZ effect. There are several techniques, 
based on semi-analytic computations and numerical simulations, 
that can be used to predict the SZ signal from a population of clusters.
The most recent numerical simulations (see e.g. \cite{white}), 
assuming cosmological parameters consistent with 
the primary anisotropies and other observables, predict, in general, 
an amplitude for the SZ
power spectrum at small angular scales which is roughly an order of 
magnitude lower than the observed by recent two experiments.

As discussed in the literature \cite{bond,selkom}, the key parameter in 
fixing the amplitude of the SZ signal is the value of the rms mass 
fluctuations on spheres of 8 Mpc h$^{-1}$ ($\sigma_8$).
Due to the highly non linear behaviour, the SZ thermal contribution 
(see Fig.2 left panel) is strongly dependent on $\sigma_8$ with 
$C_{\ell}^{SZ}\sim\sigma_8^7$. As we can see in Figure~\ref{fig:sz}, 
right panel, the power spectrum associated with the SZ effect 
(calculated following the  techniques described in \cite{Coo00,Coo01a}) 
is in agreement with the new CBI and BIMA data only for values of 
$\sigma_8 \ge 1$. This is consistent with the value of
$\sigma_8=1.05\pm0.15$ at $95 \%$ c.l. recently found by
Komatsu and Seljak \cite{selkom} in a detail analysis and with the numerical 
results obtained by both Bond et al. \cite{bond}, White et al.
\cite{white} and in Ref. \cite{yen} ($\sigma_8=1.0-1.2$).

A value of $\sigma_8 \ge1$, while in agreement
with cosmic shear data (see e.g. \cite{refregier}) and the
'old' cluster normalization \cite{pierpaoli}, is in 
disagreement with new cluster abundances results 
\cite{viana,sels8} and the cluster mass function from 
early SDSS data \cite{bahcall} when combined with the latest
CMB+Supernovae type Ia analysis $\Omega_M=0.31\pm0.05$
\cite{sievers} (see Fig.1). 
Recent joint analysis from the mass function and power spectrum obtained
from the REFLEX X-ray survey \cite{guzzo}, 
confirm these values with $\Omega_M=0.34 \pm 0.03$ and
$\sigma_8=0.71\pm0.04$.

A value of $\sigma_8$ higher than one is also 
inconsistent with the $\sigma_8\sim0.75\pm0.05$ results obtained by
combined analysis of CMB and galaxy clustering data from the recent 2dF
survey \cite{lahav,melk} 
A possible way to solve the tension with the CMB value 
is to advocate an early reionization of the intergalactic medium, 
such that the optical depth of the universe is $\tau_c\sim0.3$
\cite{bond} or a red-shift of reionization
$z_{re}\sim 20-30$. However, this is not in agreement with
the observed evolution of the Ly-alpha transmitted flux
in the spectra of four highest redshift quasars discovered
by the Sloan survey, which suggests $z_{re}=6\pm1$ \cite{gnedin} 
that points to $\tau_c\sim0.05$ (see e.g. \cite{bruscoli}).

Even if is fair to say that most of the above measurements can 
be strongly affected by systematics, it appears
that the cosmological informations obtained
from the {\it secondary} anisotropies under the assumption of a
pure SZ component are in tension with the same
informations obtained by the {\it primary} anisotropies
when combined with other cosmological observables.

A solution to this problem, however, is to postulate an extra 
contribution to the small scale anisotropies 
from some mechanism different from SZ.
Additional contributions at these small scales like those from
kinetic SZ effect, Ostriker-Vishniac effect \cite{OstVis86} 
and from the patchy, or inhomogeneous, reionization \cite{Aghetal96}. 
are generally expected much lower than the thermal SZ contribution 
\footnote{We refer the reader to Ref. \cite{Coo02} and 
\cite{Aghetal02} for a detailed discussion of these contributions.}.

\section{Beyond SZ: Contributions to Small Scale Power}

In addition to such processes,
the anisotropies at small angular scales can also be affected by 
the presence of foregrounds. At frequencies relevant to recent small 
scale anisotropy results at 30 GHz, a dominant contribution may come from
extra-galactic radio point sources. 
The possible contamination involving radio point sources has been 
investigated in \cite{mason} with the conclusion that the contribution 
from point sources should not be significant.
It has been recently noted \cite{Hol02}, however, that
if the excess power at small angular scales is to be explained 
by the SZ effect, then radio point sources at the observing
frequencies of CBI and BIMA, 30 GHz, must be a surprisingly weak
contaminant.

In this {\it paper} we want to point out that an extra 
contribution to temperature anisotropies at small angular scales 
can arise also in non-standard models of inflation. 
A strong possibility, for example, is the one that has recently been discussed 
in detail \cite{louise} involving the existence of primordial voids in the 
early universe\footnote{For the purpose of this discussion, 
note that we define early universe to be the era around, and before, 
recombination $(z \sim 1000)$, while the local universe is the era
after reionization $(z \lesssim 10)$.} \cite{Ameetal98,Saketal99,louise}. 
There is an important contribution from these voids to temperature 
anisotropies. 
Similar to the Sachs-Wolfe (SW; \cite{SacWol67}) effect associated with 
the dark matter potential at the last scattering
surface, these primordial voids also generate a SW contribution.
The angular scale for this latter SW temperature fluctuations are 
consistent with the projected size of the void at the last scattering surface.
Note that voids which are fully embedded in the primordial 
photon-baryon fluid do not generate a new anisotropy contribution via
the SW effect. 

Following the discussion presented in Ref. \cite{Saketal99} and
\cite{louise},  for illustration purposes,
we calculate the SW contribution associated with voids
using parameters which are consistent with voids observed via 
red-shift surveys of the present day and a volume fraction, again,
consistent with such observations \cite{HoyVog02}. 
Between the last scattering surface and today, 
these voids also contribute to additional temperature anisotropies 
through frequency shifts, mainly the Rees-Sciama effect (RS; \cite{ReeSci68})
and effects such as gravitational lensing. 
For parameters on voids consistent with
current observations, these latter low red-shift contributions are however 
smaller and can be ignored \cite{Saketal99}.

Another possibility, recently investigated by various authors
(see e.g. \cite{features}) but in different contexts, is the presence
of a feature in the primordial spectrum of fluctuations, as expected,
for example, in inflationary models with broken scale invariance.
In Fig.1 we present the possible contribution to the small scale
anisotropies from a feature in the spectrum modeled as gaussian
centered at $k \sim 0.20 [h Mpc^-1]$ with dispersion 
$\Delta k \sim 0.03 [h Mpc^-1]$ and amplitude $A \sim 5$.
While this feature is not fully ruled out by
results on the matter power spectrum from recent galaxy red-shift surveys 
such as the  2dF (see e.g. \cite{teg2df}), its width is still dangerously
close to the observed spectral resolution around these scales. 
Though models involving primordial features only produce excess 
power over a limited range in multipolar space,
in this case, only out to a $l \sim 3000$, the current 
small scale anisotropy observations are also limited in
the coverage of the power spectrum complicating any identification of 
a feature. However, as we can see from Fig.2, the BIMA
result is already strongly constraining this hypothesis.

In addition to all these possible contributions arising from modifications
to the standard inflationary scenario, we also note that there 
may be additional possibilities to generate small scale anisotropies 
at the last scattering via modifications to recombination \cite{Whi01}.

Given distinct possibilities for small scale temperature anisotropies,
involving the large scale structure after reionization and the last
scattering surface, and giving the difficulties in explaining the 
CBI effect as solely due to SZ, an important question is
how to distinguish between them when interpreting any detection of power
at smaller angular scales in current and future experiments.

In the rest of the discussion, we will therefore consider how likely
we can distinguish between the two scenarios involving an early universe 
contribution or a local redshift contribution. 
For the purpose of this discussion we will assume potential low redshift 
contributions involving SZ and radio point sources.

\section{Early Universe or Local Structures?}

Note that a fundamental aspect related to the SZ contribution is its 
distinct frequency dependence. This spectrum can be utilized to separate its
contribution from the dominant anisotropies associated with primary
fluctuations at large scale and other thermal fluctuations, such as due 
to the SZ kinetic effects associated with the peculiar motions of clusters, 
as well as any void or non-standard contribution, at small angular scales 
\cite{Cooetal00}.
The current anisotropy observations at small scales, unfortunately, are
limited at most to a single frequency and this
limitation is unlikely to be improved significantly till the advent
of next generation of experiments.
In the case of CBI and BIMA, observations are limited to 30 GHz at the
RJ part of the spectrum.
We expect observations, at higher frequencies,
such as at 150 GHz by the ACBAR instrument to see a contribution which is
lower than at 30 GHz, by a factor of $\sim$ 0.22. 
Observations at and above the SZ null frequency of 217 GHz
are clearly desirable since a contribution from radio point sources 
is also expected to be decreasing in frequency while an early-universe 
contribution is frequency independent.

An additional aspect of the SZ effect is its non-Gaussianity. It is now well 
established that contributions to the SZ effect
primarily comes from massive galaxy clusters which are rare. Such a mass 
dependance makes the SZ effect highly non-linear.  As illustrated
in figure~\ref{fig:sz}, for example, 
the SZ effect varies by a factor of $\sim$ 2 when the normalization 
of the matter power spectrum is
changed by $\sim$ 10\%.  Another aspect of this non linearity is 
the increase in SZ variance when compared to the expected
Gaussian variance contribution. 
In addition to SZ, point source contribution may also be non-Gaussian, 
especially, if point sources trace the non-linear large scale structure 
at low redshifts.
Thus, non-Gaussian aspects are potentially common to all low red-shift 
contributions since they all tend to trace the local universe non-linear 
structures which are non-Gaussian distributed.

The non-Gaussianity alone, however, may not distinguish the 
nature of small scale anisotropy power as observed in current experiments 
as the non-standard modifications 
to the last scattering involving  primordial voids also generate a highly 
non-Gaussian anisotropy 
contribution \cite{Cooetal02}.  
A reliable approach to distinguish between a local, 
whether SZ or point sources,
and an early contribution, voids and bumps in the primordial power spectrum, 
is to consider a combined study involving the large scale structure  and CMB.
Here, we suggest a cross-correlation of the CMB anisotropy data with 
a map of the large scale structure.

The correlation between CMB, mainly the best COBE DMR map, and large scale 
structure has already been considered to understand the extent to which
the ISW effect contributes at large angular scales \cite{CriTur96}. 
As discussed in \cite{Coo02a}, this correlation, however, is dominated by
the large cosmic variance at low multipoles corresponding to angular 
scales with tens of degrees on the sky. In the case of small
scales anisotropies, the extent to which the correlation 
can be detected will be determined primarily by the
instrumental noise contribution. For the purpose of this discussion, 
we introduce the correlation coefficient between, say, large scale structure 
contributions (SZ or point sources) and a tracer of the large scale 
structure as

\begin{equation}
{\rm Corr}(\lss,i)_l  = \frac{C^{\lss-i}_l}{\sqrt{C^\lss_l C^i_l}} \, ,
\end{equation}
where $C^{\lss-i}_l$ is the cross power spectrum between large scale 
structure contributions and the tracer field. 
We write this as $C^{\lss-i}_l=C^{\sz-i}_l + C^{\ps-i}_l$, 
where $C^{\sz-i}_l$ and $C^{\ps-i}_l$ are the cross
power spectra between SZ and the tracer field, and, point sources (PS) 
and the tracer field respectively.
The cross correlation of SZ and the tracer field is described in 
Ref. \cite{Coo00} following the
so-called halo model approach to large scale structure \cite{CooShe02}; we 
note that a similar approach can be considered for $C^{\ps-i}_l$.  
Given the poor knowledge of the
statistical distribution of radio-galaxies, which could represent a 
different population from those observed in the
optical, and the wide range of observed spectral energy distributions 
due to different synchrotron cut-offs, however, 
make the problem of predicting radio source
clustering properties in a given band, as well as their 
correlation with other LSS tracers, a difficult,
though interesting, task.
We note that certain issues related to point sources are 
discussed in \cite{Hol02} to which we refer the reader for further details.

Note that the associated signal-to-noise for the detection of the 
cross-correlation is

\begin{equation}
\left(\frac{\rm S}{\rm N}\right)^2 = f_{\rm sky} \sum_l \frac{(2l+1){\rm Corr}^2(\lss,i)_l}{{\rm Corr}^2(\lss,i)_l+
\left(1+\frac{N_l^\lss}{C^\lss}\right)\left(1+\frac{N_l^i}{C^i_l}\right)} \, ,
\end{equation}
where $C^\lss=C_l^{\rm SZ}+C_l^{\rm PS}+C_l^{\rm SZ-PS}$, $N_l^\lss (=C_l^{\rm CMB}+C_l^{\rm LSS}+C_l^{\rm noise})$ and 
$N_l^i$ are the total power spectrum contributions from large scale structure, 
noise contributions associated with small scale fluctuations and the power  spectrum of the $i$ tracer field, respectively.
Here, $C_l^{\rm SZ-PS}$ is the cross-power spectrum between SZ effect and contaminant point sources; the
latter can again be calculated following models of Ref. \cite{Coo02}. If $i$th field is galaxies (which may also
contribute to radio background), then we note that $C_l^{\rm SZ-i} \propto C_l^{\rm SZ-PS}$. Note that
at small scales, $C_l^{\rm CMB} \ll C_l^{\rm LSS}$.

In figure~\ref{fig:corr}, we show the correlation coefficient of SZ and 
tracers of the large scale structure
involving weak lensing convergence and the galaxy distribution. 
In the near future, a cross-correlation study of small scale anisotropies
and a map of the galaxy distribution, such as from the Sloan survey, looks 
promising and should be considered.  
To calculate the expected signal-to-noise ratio for a detection of the 
correlation signal, we assume a noise contribution
to the tracer field involving the shot-noise contribution arising from the 
finite number of galaxies and assume no point source  contribution to small 
scale anisotropy signal. The latter is a safe assumption given that
point source contributions have been found to be subdominant by various 
monitoring and analysis techniques considered by, at least, the CBI group. 
For a survey of 10 degrees$^2$ and no noise contributions, we 
estimate the signal-to-noise ratio of order $\sim$ 100 
while this drops to $\sim$ 10 when reasonable noise contributions are 
considered to both the temperature and galaxy
tracer field. For example, the galaxy shot-noise considered in this involve 
a surface density of $10^9$ sr$^{-1}$, which is the density of galaxies 
down to a R band magnitude of 25 \cite{Smaetal95}. 
We note that the presence of point sources will likely increase this 
signal-to-noise estimate considerably since point sources may correlated 
between with tracers of the large scale structure, such as the galaxy 
distribution field. 
This increase clearly motivates attempts for such a combined study.

Thus, the extent to which small scale anisotropies correlate with large scale 
structure by it self does not determine the
nature of the temperature fluctuation. Though it establishes that the local 
universe is partly responsible, it does not
necessarily mean that SZ is the dominant contribution since fluctuations may 
contain a point source contribution.
If one is interested in distinguishing between SZ and point sources, it is 
necessary to consider aspects beyond
simple cross-correlations.

In general, the extent to which  SZ thermal effect contributes at small 
angular scales can be established more 
reliably based on its spectral dependence relative to thermal CMB than any 
correlation associated with the large
scale structure alone. The SZ frequency spectrum is unique and is unlikely 
to be mimicked by other sources of contributions.
To determine the SZ contribution and to separate it from other thermal 
fluctuations at small
angular scales, the future experiments should be equipped with 
multifrequency capabilities that expands from
low RJ frequencies to $\sim$ 300 GHz; the low frequencies help 
determine the radio source contribution while high frequencies determine
the confusion from mid-infrared/submm point sources. 
The observations at SZ null of 217 GHz determine the extent to which
other thermal fluctuations are significant as a source of 
small scale anisotropies. 

Though to a certain extent current and upcoming
detection of small scale power may be important as a first detection, 
dedicated small angular scale
experiments with multifrequency coverage is clearly needed to fully 
understand the nature of fluctuations at these small scales.
Just as anisotropy studies at degree angular scales involving the 
acoustic peak structure have been successful as a strong
probe of cosmology, the small-scale anisotropies open 
the window to understand both the large scale structure and important subtle
effects involving the last scattering surface.

\acknowledgments
We wish to thank Martin Kunz, Louise Griffiths and Joe Silk for helpful 
comments. This work was supported at Caltech by the Sherman Fairchild 
foundation and the DoE grant DE-FG03-92-ER40701. AM is supported by PPARC.

\end{document}